\documentclass[reprint,amsmath,amssymb,aps,prl]{revtex4-2}
\usepackage[pdftex]{graphicx}
\usepackage{dcolumn}
\usepackage{txfonts}
\usepackage{bm}
\usepackage{color}
\usepackage{hyperref}

\usepackage{pdfpages}
\makeatletter
\AtBeginDocument{\let\LS@rot\@undefined}
\makeatother

\DeclareMathOperator{\sgn}{sgn}

\begin{document}

\title{%
Low-Rank Combinatorial Optimization and
Statistical Learning by Spatial Photonic Ising Machine}
\author{Hiroshi Yamashita}
\author{Ken-ichi Okubo}
\author{Suguru Shimomura}
\author{Yusuke Ogura}
\author{Jun Tanida}
\author{Hideyuki Suzuki}
\email{hideyuki@ist.osaka-u.ac.jp}
\affiliation{%
  Graduate School of Information Science and Technology,
  Osaka University, Osaka 565--0871, Japan}

\begin{abstract}
The spatial photonic Ising machine (SPIM)
[D. Pierangeli et al., Phys. Rev. Lett. \textbf{122}, 213902 (2019)]
is a promising optical architecture
utilizing spatial light modulation for solving
large-scale combinatorial optimization problems efficiently.
The primitive version of the SPIM, however, can accommodate
Ising problems with only rank-one interaction matrices.
In this Letter, we propose a new computing model for
the SPIM that can accommodate any Ising problem
without changing its optical implementation.
The proposed model
is particularly efficient for Ising problems with low-rank interaction matrices,
such as knapsack problems.
Moreover, it acquires the learning ability of Boltzmann machines.
We demonstrate that learning, classification, and sampling of the MNIST handwritten digit images
are achieved efficiently using the model with low-rank interactions.
Thus, the proposed model exhibits higher practical applicability
to various problems of combinatorial optimization and statistical learning,
without losing the scalability inherent in the SPIM architecture.
\end{abstract}

\maketitle

\textit{Introduction.}%
---%
As the recent development of machine intelligence technologies relies
largely on massive computational power for optimization and learning,
there is a growing demand for high-speed, large-scale,
and energy-efficient computation
to deal with increasingly complex real-world problems.
A possible approach to meet this demand is to
adopt unconventional, problem-specific computing technologies,
without relying on the conventional von Neumann architecture.

Ising machines are dedicated hardware solvers for
combinatorial optimization problems formulated as Ising problems,
designed to find the (approximate) ground states of the corresponding Ising models
\cite{Ackley1985,Korst1989}.
Many important combinatorial optimization problems can be formulated
as Ising problems \cite{Korst1989,Lucas2014},
thus leading to numerous studies
\cite{Mohseni2022,Johnson2011,Yamaoka2015,Inagaki2016,Leleu2019,Goto2019,Tsukamoto2017,Okuyama2019,Wang2019,Pierangeli2019,Pierangeli2020,Bohm2022}
on implementing Ising machines
using various physical devices and dynamics.

The spatial photonic Ising machine (SPIM) \cite{Pierangeli2019,Pierangeli2020}
is a promising optical architecture utilizing spatial light modulation
for solving large-scale Ising problems efficiently.
The SPIM accelerates annealing computation
by optically computing the Ising Hamiltonian with all-to-all interactions
in constant time,
independent of the number of variables.
Its outstanding performance has been demonstrated
for problems with more than ten thousand variables \cite{Prabhakar2023}.

Despite its superior scalability, the primitive version of
the SPIM can accommodate only a limited class of Ising problems
with rank-one interaction matrices.
Although subsequent studies \cite{Sun2022,Luo2023} multiplexed
the SPIM to handle broader classes of Ising problems,
the scalability becomes degraded instead.
Thus, a breakthrough is still required for the SPIM to attain the applicability
to various real-world problems without losing its scalability.

In this Letter, we propose a multicomponent computing model for the SPIM
to circumvent the limitation and accommodate higher-rank interaction matrices
without changing its optical implementation.
The proposed model is capable of handling any Ising problem,
and is particularly efficient for problems with low-rank interactions.
We demonstrate its efficient applicability to knapsack problems
by formulating them as Ising problems with rank two.

Moreover, we show that the proposed model acquires
the learning ability of Boltzmann machines \cite{Ackley1985}.
With full-rank interactions, it has
the expressive power equivalent to the ordinary Boltzmann machine;
however, the model with low-rank interactions is efficient and
can be sufficient for inferences from real-world data,
as typically assumed in low-rank modeling.
We demonstrate that learning, classification, and sampling
of the MNIST handwritten digit images \cite{MNIST} are achieved efficiently
with low-rank interactions.
Notably, we observe that the newly derived learning rule
naturally performs low-rank learning of the digit images,
whereas low-rank constraints are not explicitly imposed.

Thus, we report here that the proposed model
exhibits higher practical applicability
to various problems of combinatorial optimization and statistical learning,
without losing the scalability inherent in the SPIM architecture.
Although our contribution in this Letter is the computing model
that theoretically works with any existing SPIM implementation,
we also present the results of proof-of-concept optical experiments.

\begin{figure}[tb]
\centering
\includegraphics[width=80mm]{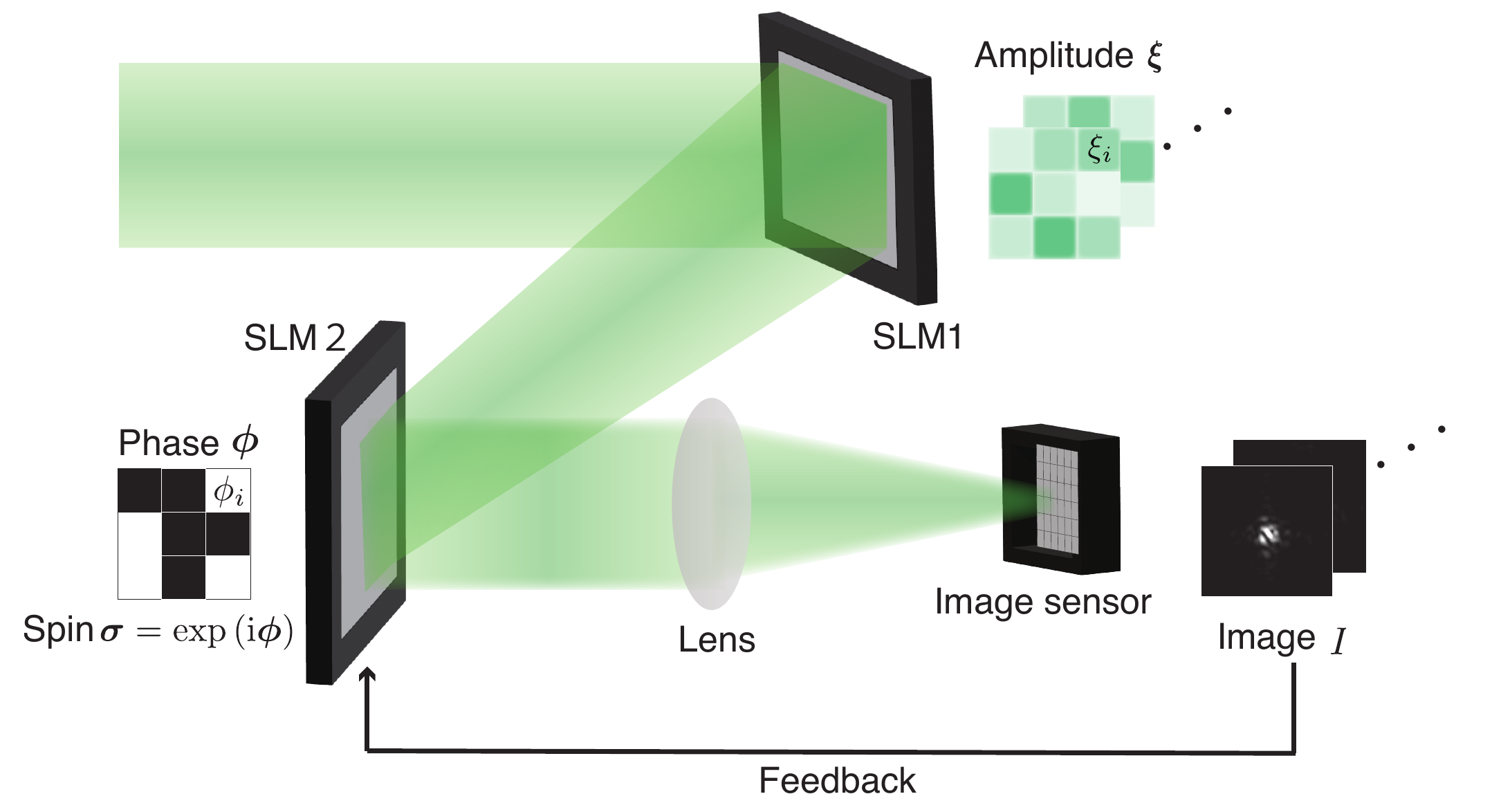}
\caption{Schematic of the SPIM architecture.
The laser beam is amplitude-modulated and phase-modulated
by spatial light modulators SLM1 and SLM2,
which encode $\bm\xi$ and $\bm\sigma$, respectively,
and detected by an image sensor.
The Ising Hamiltonian is obtained from the detected image $I$.}
\label{fig:schematic}
\end{figure}

\textit{Optical computation of Ising Hamiltonian.}%
---%
The SPIM \cite{Pierangeli2019,Pierangeli2020}
computes the Ising Hamiltonian optically from the phase-modulated image
of an amplitude-modulated laser beam (Fig.~\ref{fig:schematic}).
Light incident on the $i$th site of the spatial light modulator (SLM)
with an amplitude $\xi_i$
is phase-modulated by $\sigma_i=\exp(\mathrm{i}\phi_i)=\pm 1$,
which represents the $i$th Ising spin,
and detected by an image sensor.
In the primitive version of the SPIM,
the detected image $I$ is compared with the point-like target image $I_\text{T}$
to obtain the Ising Hamiltonian in the following form:
\begin{equation}
H(\bm\sigma) \propto \sum_{i,j}\xi_i\xi_j\sigma_i\sigma_j
= \bm\sigma^\top\bm\xi\bm\xi^\top\bm\sigma,
\label{eq:spim}
\end{equation}
where $\bm\xi=(\xi_1,\ldots,\xi_N)^\top$
and $\bm\sigma=(\sigma_1,\ldots,\sigma_N)^\top$.
Notably, the computation is performed in constant time,
independent of the number of spins $N$,
involving all-to-all interactions among the spins.
However, compared with the ordinary (quadratic) Ising Hamiltonian
$H(\bm\sigma)=-\frac{1}{2} \bm\sigma^\top J\bm\sigma$,
the interaction matrix $J$ is limited to the form
$J\propto\bm\xi\bm\xi^\top$.
Thus, the primitive SPIM can accommodate only real symmetric matrices with rank one
as the interaction matrix.
The Ising spin system with this type of Hamiltonian is known as
the Mattis model \cite{Mattis1976}.

Here we propose a multicomponent computing model for the SPIM architecture
to improve the expressive power of the interaction matrix.
We formulate the Hamiltonian
as a linear combination of Eq.~(\ref{eq:spim}) as follows:
\begin{equation}
H(\bm\sigma)=-\frac{1}{2}\sum_{k=1}^K\lambda_k\sum_{i,j}\xi_{i,k}\xi_{j,k}\sigma_i\sigma_j
= -\frac{1}{2}\bm\sigma^\top
   \left(\sum_{k=1}^K\lambda_k\bm\xi_k\bm\xi_k^\top\bm\right)
   \bm\sigma,
\label{eq:multicomponent}
\end{equation}
where $K$ denotes the number of components,
and $\lambda_k$ and $\bm\xi_k=(\xi_{1,k},\ldots,\xi_{N,k})^\top$ are
the weight and amplitude parameters of the $k$th component, respectively.
The energy value of the Hamiltonian can be obtained by calculating
the weighted sum from images acquired $K$ times
with different amplitudes $\bm\xi_k$.
Now the interaction matrix
$J=\sum_k\lambda_k\bm\xi_k\bm\xi_k^\top$ 
can represent any real symmetric matrix with rank not greater than $K$.
Therefore, if $K$ is increased to $N$,
any Ising Hamiltonian can be computed.
Although the computation time increases linearly to $K$,
it does not depend directly on $N$,
inheriting the scalability of the underlying SPIM architecture.

\textit{Combinatorial optimization with the multicomponent model.}%
---%
To solve a combinatorial optimization problem using an Ising machine,
we formulate it as an Ising problem,
which is to find $\bm\sigma\in\{+1,-1\}^N$ that minimizes
the Ising Hamiltonian $H(\bm\sigma)=-\frac{1}{2}\bm\sigma^\top J \bm\sigma$.
For simplicity, the linear (bias) term is omitted here
because introducing an additional spin fixed to $+1$ suffices.

The Hamiltonian of the primitive SPIM, with rank $K=1$,
is $H(\bm\sigma)=-\frac{\lambda}{2}\left(\bm\xi^\top\bm\sigma\right)^2$.
When $\lambda>0$, it has trivial, two symmetric global minima $\bm\sigma=\pm\sgn\bm\xi$.
When $\lambda<0$, minimizing $H(\bm\sigma)$
reduces to a number partitioning problem
\cite{Ferreira1998,Mertens1998,Mertens2001,Pierangeli2021,Huang2021,Prabhakar2023},
which is to find the partition of numbers $\xi_1,\dots,\xi_N$
into two subsets that minimizes the difference of the sums in the two subsets
$\left\lvert\sum_i \xi_i\sigma_i\right\rvert=\left\lvert\bm\xi^\top\bm\sigma\right\rvert$.
Thus, the primitive SPIM can essentially handle
only the class of number partitioning problems.
Although this class is theoretically NP-hard \cite{Pedroso2010},
it is practically insufficient to be used
for solving Ising formulations of various combinatorial optimization problems.

However, we can circumvent the limitation without changing the optical implementation by introducing
the proposed multicomponent model, which is capable of handling any Ising problem.
Particularly, it is efficient
for Ising problems with low-rank interactions
because the computation time depends linearly on rank $K$.

The spin configuration $\bm\sigma$ is updated according to
energy values $H(\bm\sigma)$.
To solve an Ising problem, typically we employ simulated annealing \cite{Kirkpatrick1983};
a sample sequence of $\bm\sigma$,
generated by a Markov-chain Monte Carlo (MCMC) method
from the Gibbs distribution
$P(\bm\sigma)\propto\exp \left[-H(\bm\sigma)/T\right]$,
is expected to converge to an approximate ground state
as the system temperature $T$ gradually decreases.

\textit{Application to knapsack problems.}%
---%
To demonstrate the applicability of the multicomponent model
to a broader class of combinatorial optimization problems,
we apply it to the 0-1 knapsack problem with integer weights,
which can be formulated as Ising problems with rank $K=2$
and hence cannot be handled by the primitive SPIM.

The knapsack problem is a well-known problem
to find the subset of given items that maximizes the total value
satisfying a predefined total weight limit.
More specifically, given the value $v_i$ and the weight $w_i$ of
the $i$th item for $i=1,2,\ldots,n$ and the weight limit $W$,
the 0-1 knapsack problem is expressed as follows:
\begin{align}
&\text{maximize}\quad \sum_{i=1}^nv_i x_i\\
&\text{subject to}\quad \sum_{i=1}^n w_i x_i \le W,
\quad \bm{x}=(x_1,\dots,x_n)\in\{0,1\}^n.
\end{align}
Under the assumption of integer weights, the knapsack problem
reduces to minimizing
\begin{equation}
H(\bm{x},\bm{y})=
  A\left(\sum_{i=1}^nw_ix_i + \sum_{i=1}^m2^{i-1}y_i - W\right)^2
  -B\left(\sum_{i=1}^n v_ix_i\right)^2,
\label{eq:knapsack}
\end{equation}
where auxiliary variables $\bm{y}=(y_1,\dots,y_m)\in \{0,1\}^m$
are introduced using a log trick \cite{Lucas2014}.
This can be rewritten
in the multicomponent form (\ref{eq:multicomponent})
with size $N=n+m+1$ and rank $K=2$ as follows:
\begin{align}
&\lambda_1 = -\frac{A}{2}, \quad \lambda_2 = +\frac{B}{2},\\
&\bm\xi_1=\bigl(w_1,\ldots,w_n,2^0,\ldots,2^{m-1},\textstyle\sum_iw_i+2^m-1-2W\bigr)^\top,\\
&\bm\xi_2=\bigl(v_1,\ldots,v_n,0,\ldots,0,\textstyle\sum_iv_i\bigr)^\top,\\
&\bm\sigma = \left(2x_1-1,\ldots,2x_n-1,2y_1-1,\ldots,2y_m-1,1\right)^\top.
\end{align}

\begin{figure}[tb]
\centering
\includegraphics{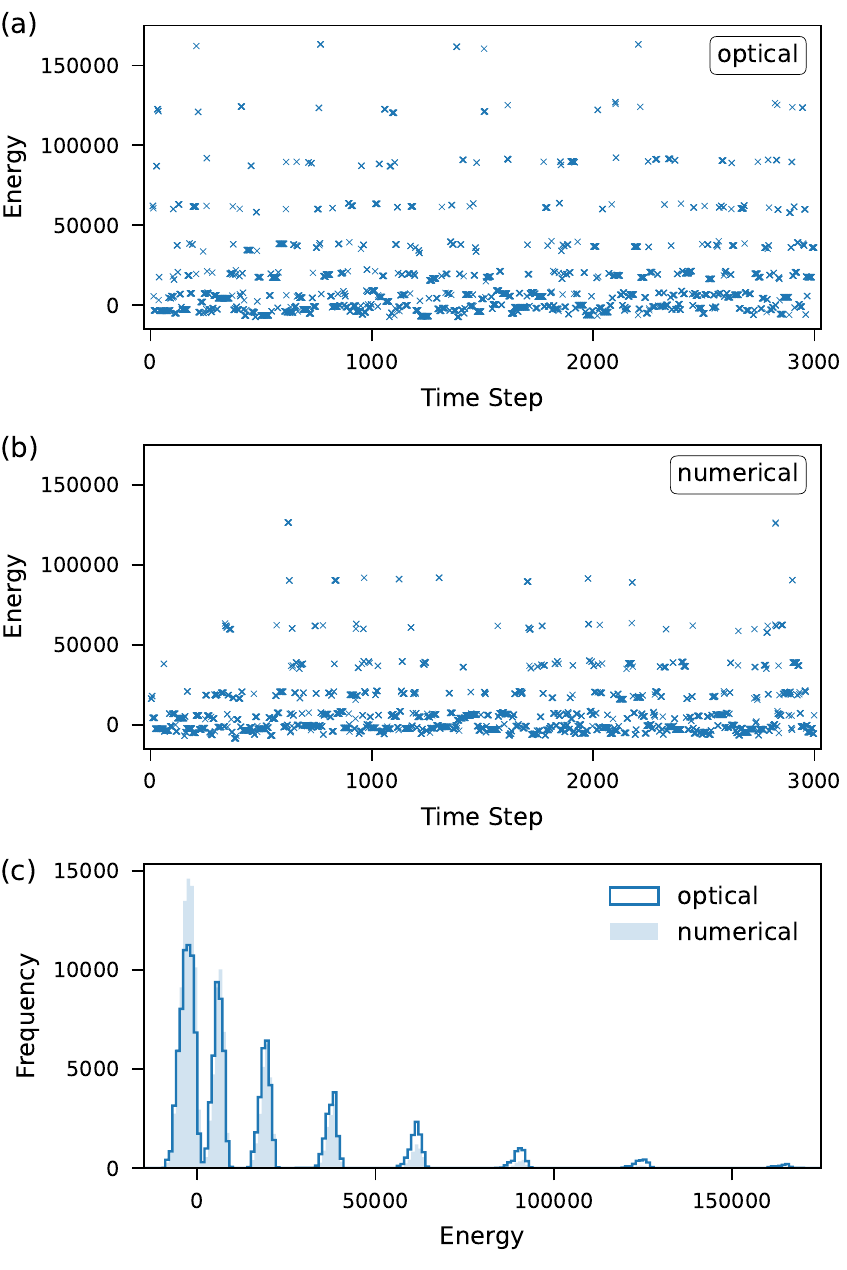}
\caption{Sampling behavior of the multicomponent model for a knapsack problem.
Typical time evolutions of energy values
of the spin configurations sampled
from (a) optical and (b) numerical experiments.
Several samples with higher energy values are not shown.
(c) Histograms of the energy values of $3000\times 50$ samples
observed from each experiment with bin width 1000.}
\label{fig:knapsack}
\end{figure}

We conducted a proof-of-concept experiment \cite{SM}
for a knapsack problem with $n=13$ items \cite{Pisinger1999}.
The spin sequences were sampled both optically and numerically
at a moderately low, constant temperature.
Fig.~\ref{fig:knapsack} shows
that the multicomponent SPIM generates samples essentially
according to the Gibbs distribution.
The typical time evolution of the energy values $H(\bm\sigma)$ of spins
observed in the optical experiment (Fig.~\ref{fig:knapsack}(a))
resembles that of the numerical experiment (Fig.~\ref{fig:knapsack}(b)).
The histogram of energy values sampled from the optical experiment
(Fig.~\ref{fig:knapsack}(c)) shows that it generates
many low-energy samples around $H(\bm\sigma)\approx 0$,
constituting the distribution
with peaks at the same values as those in the numerical experiment.

A closer look at these results indicates that
the temperature of the Gibbs distribution
was slightly higher in the optical experiment
due to the noise in the optical system.
Although the physical noise can be utilized as a source of randomness \cite{Pierangeli2020NP},
we simply executed the Metropolis algorithm
adhering to the obtained Hamiltonian values for clarity of results.
To facilitate the MCMC process to jump over energy barriers,
multiple-spin flips were performed, taking the advantage of
direct energy computation of the SPIM.

The optimal solution to the knapsack problem was obtained 304 times
out of the 150000 samples observed in the optical experiment,
with a ratio considerably higher than the probability $2^{-13}$ of random sampling.
This result confirms that the spin states with lower energy values
were sampled frequently according to the Gibbs distribution
$P(\bm\sigma)\propto\exp \left[-H(\bm\sigma)/T\right]$.

Overall, we demonstrated that the multicomponent model with rank $K=2$ works as expected
with the Ising Hamiltonian for the knapsack problem
in both the numerical and optical experiments.
These results indicate that the proposed model can efficiently handle
Ising problems with low-rank interactions.

\textit{Statistical learning with the multicomponent model.}%
---%
In the field of machine learning, the Ising model is commonly referred to
as the Boltzmann machine, which can be viewed as a generative
neural network model composed of stochastic elements \cite{Goodfellow2016}.
It has been applied not only for
solving combinatorial optimization problems \cite{Korst1989}
but also, more importantly, for statistical machine learning.
The restricted Boltzmann machine (RBM) \cite{Smolensky1986,Hinton2002,Larochelle2008}
and deep Boltzmann machine (DBM) \cite{Salakhutdinov2009,Salakhutdinov2012}
are well-known subclasses that
have contributed to the recent development of deep learning.

With the increased expressive power,
the multicomponent model acquires the learning ability
applicable to real-world data.
If rank $K$ is increased to $N$,
it becomes equivalent to the ordinary Boltzmann machine;
however, it is efficient with low-rank interactions
both in terms of the computation time and the number of parameters.

To train the model $P(\bm\sigma)\propto\exp \left[-H(\bm\sigma)\right]$,
we perform the gradient ascent on the log-likelihood $\log L$
given the data distribution, according to the gradients \cite{SM}
\begin{align}
  \frac{\partial}{\partial \lambda_k}\log L
  &= \frac{1}{2} \bm\xi_k^\top\left(
      \langle\bm\sigma\bm\sigma^\top \rangle_\text{data}
    - \langle\bm\sigma\bm\sigma^\top \rangle_\text{model}
    \right)\bm\xi_k, \label{eq:gradient-lambda}\\
  \frac{\partial}{\partial \bm\xi_k}\log L
  &= \lambda_k\left(
      \langle \bm\sigma\bm\sigma^\top \rangle_\text{data}
    - \langle \bm\sigma\bm\sigma^\top \rangle_\text{model}
    \right)\bm\xi_k, \label{eq:gradient-xi}
\end{align}
where $\langle \cdot \rangle_\text{data}$
and $\langle \cdot \rangle_\text{model}$
denote the expectations over
the data and model distributions, respectively.

\textit{Learning MNIST digit images.}%
---%
To demonstrate the learning ability as a Boltzmann machine
with the low-rank efficiency,
we trained the multicomponent model \cite{SM} using the MNIST digit image data \cite{MNIST}.

\begin{figure}[tb]
\centering
\includegraphics{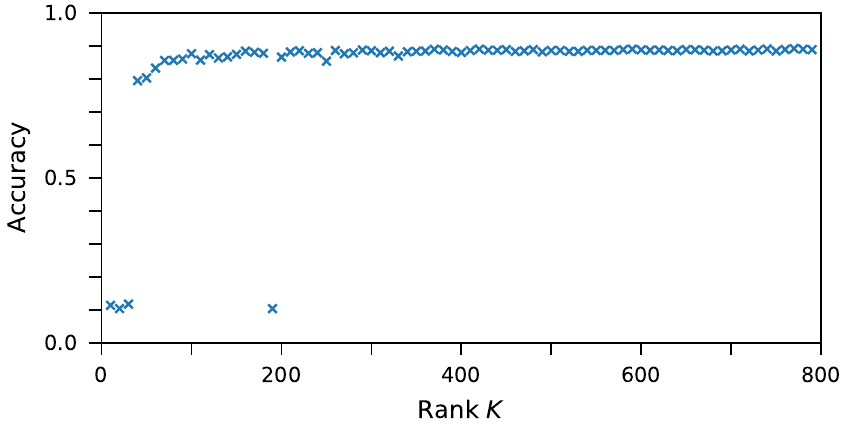}
\caption{Classification of MNIST digit images.
The classification accuracy of the trained multicomponent models
with rank $K$ taking on integer multiples of 10 is shown.}
\label{fig:classification}
\end{figure}

First, we applied it for the classification of handwritten digits
to evaluate its low-rank efficiency.
We trained the fully visible model with size $N=794$.
Fig.~\ref{fig:classification} shows the dependency of
the classification accuracy on rank $K$.
Although the accuracy drops to the chance level for $K\le 30$
owing to training failure,
the graph is almost flat for $K\ge 100$; that is, the model with rank
as low as $K=100$ exhibits a performance comparable to that of the full rank.
This numerical result clearly shows the low-rank efficiency
of the multicomponent model in learning the MNIST images.
Note that the accuracy was not as high as that of the ordinary RBM
due to the lack of hidden units.

\begin{figure}[tb]
\centering
\includegraphics{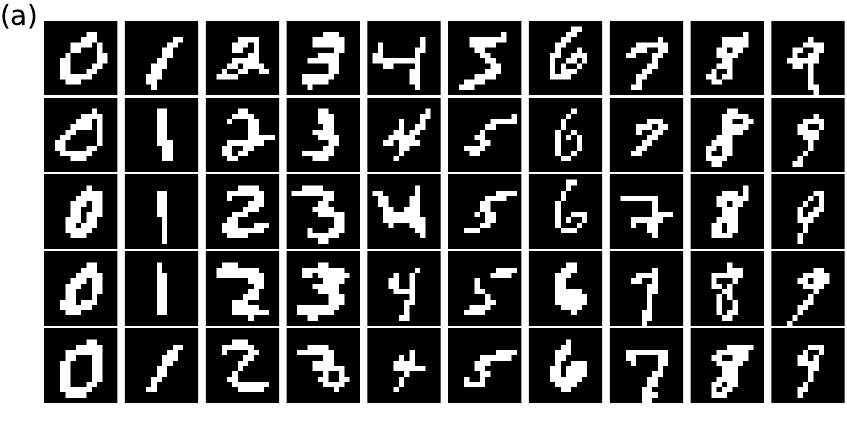}
\includegraphics{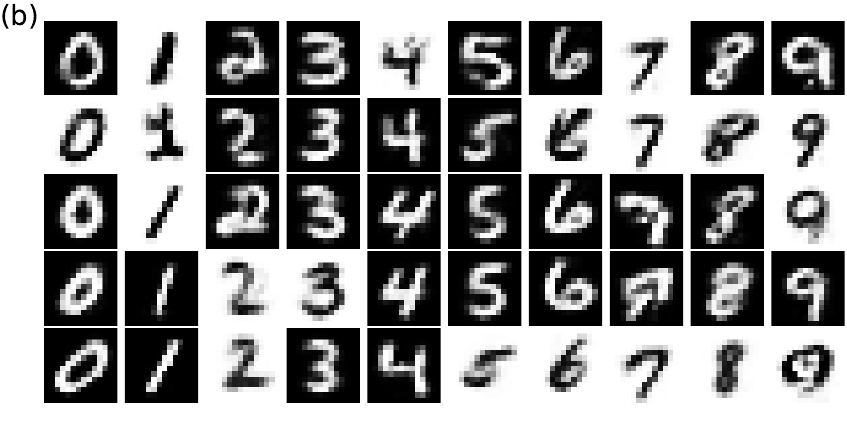}
\includegraphics{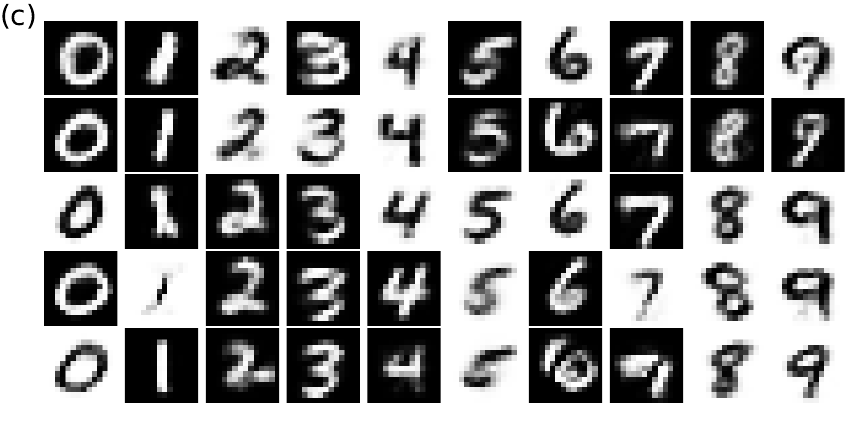}
\includegraphics{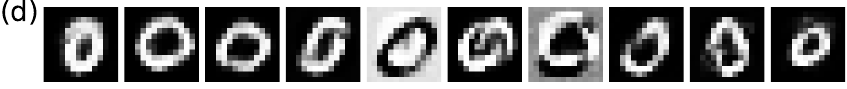}
\caption{Sampling from the multicomponent models trained with the MNIST digit images.
(a) Random samples from the training dataset.
Random samples generated from
(b) the trained models with rank $K=50$,
(c) the reduced models with only principal components, and
(d) the optical experiment of the reduced model,
after 1960 time steps from random initial spin configurations.
Each pixel in the gray-scale images represents the conditional probability
$P[\sigma_i=1\mid\bm{\sigma}_{\setminus i}]$
of each spin $\sigma_i$ given the states of other spins $\bm{\sigma}_{\setminus i}$.}
\label{fig:sampling}
\end{figure}

Next, we sampled digit images from the fully visible multicomponent models
with size $N=196$ and rank $K=50$
trained using the MNIST images of each digit
(Fig.~\ref{fig:sampling}(a)).
Random samples from the trained models (Fig.~\ref{fig:sampling}(b))
show that the digit images were successfully sampled.
Note that inverse images are sampled due to the symmetry
$H(\bm\sigma)=H(-\bm\sigma)$ of the model without bias.
The images did not degrade in random samples from the reduced model (Fig.~\ref{fig:sampling}(c))
composed only of principal components with magnitudes
$\left\lvert\lambda_k\right\rvert\left\|\bm\xi_k\right\|^2 > 0.1$.
These numerical results indicate that the reduced, low-rank model is sufficient for sampling.
Fig.~\ref{fig:sampling}(d) shows random samples obtained optically
from the reduced model for the digit ``0''.
Some samples maintain the digit shape,
while some appear to degrade, in comparison with the numerical results,
possibly due to the noise in the optical system.
Again, we did not utilize the physical noise for clarity of results.

\begin{figure}[tb]
\centering
\includegraphics{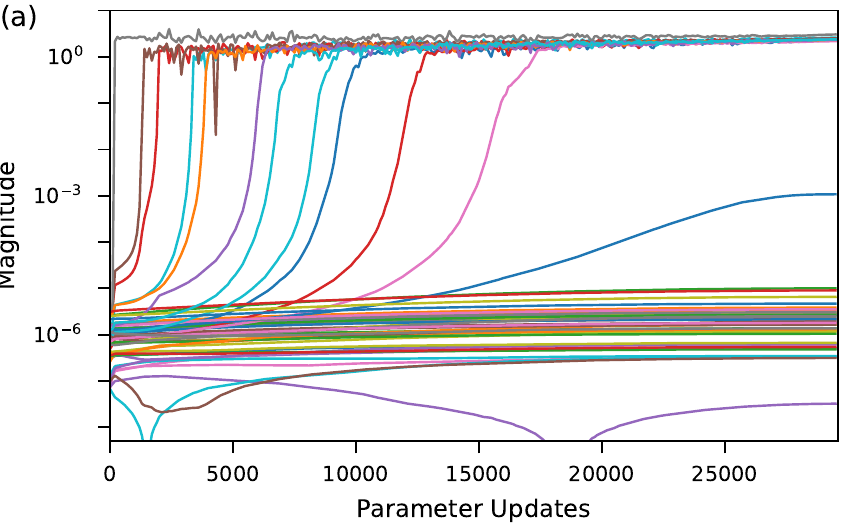}
\includegraphics{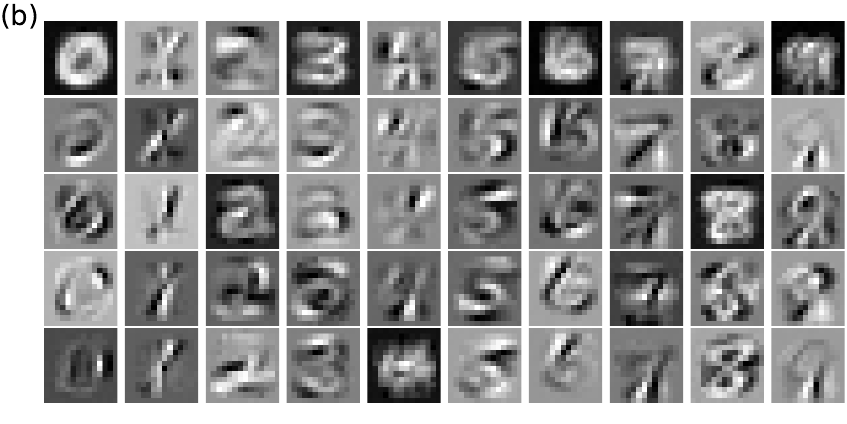}
\caption{Low-rank learning of the multicomponent model.
(a) Time evolution of the magnitude values
$\left\lvert\lambda_k\right\rvert\left\|\bm\xi_k\right\|^2$
in the learning process of the model with rank $K=50$ for the digit ``0''.
(b) Gray-scale images of principal components $\bm\xi_k$
with the five largest magnitude values for each digit.}
\label{fig:contribution}
\end{figure}

The learning behavior for the digit ``0'' is
depicted in Fig.~\ref{fig:contribution}(a).
The magnitude $\left\lvert\lambda_k\right\rvert\left\|\bm\xi_k\right\|^2$ for each $k$th
component increases one by one as the learning process progresses.
The final number of principal components is 11 out of $K=50$.
The multicomponent model appears to gradually increase its (effective) rank
as required for accuracy.
This result suggests that the gradient-ascent learning rule
naturally achieves low-rank learning.
Fig.~\ref{fig:contribution}(b) shows the gray-scale images of
the top five principal components $\bm\xi_k$,
for which $\lambda_k$ is positive.
The digit shapes are vaguely embedded in $\bm\xi_k$,
because intuitively, $\bm\sigma=\pm\sgn\bm\xi_k$
minimizes the $k$th component if $\lambda_k>0$.

Overall, both the classification and sampling results demonstrate
the low-rank efficiency of the multicomponent model
in learning MNIST digit images with the gradient-ascent learning rule.

\textit{Discussion.}%
---%
Since the multicomponent model handles lower-rank Ising problems more efficiently,
the matrix ranks can be
an index that characterizes a new aspect, to the best of our knowledge, 
for combinatorial optimization problems.
The number partitioning problem
and the 0-1 knapsack problem with integer weights
are lowest-rank examples
of combinatorial optimization problems.
It is an interesting future direction to characterize
the types of low-rank real-world problems.

The efficiency for low-rank Ising Hamiltonians
as well as inherent scalability with all-to-all interactions
is a unique feature that cannot be seen in other Ising machines.
For rank one, its outstanding performance has already been demonstrated
in solving large-scale number partitioning problems \cite{Prabhakar2023}.
Thus, the multicomponent SPIM is also expected to exhibit unique performance
for large-scale Ising problems with low-rank all-to-all interactions.

A necessity for solving low-rank Ising problems arises
when the learning is involved \cite{Kitai2020,Wilson2021,Matsumori2022}.
For example, a study on the automated design of metamaterials \cite{Kitai2020}
trains a factorization machine \cite{Rendle2010},
similar to the multicomponent model,
to find low-energy candidates for metamaterials
using a D-Wave quantum annealer.
Here, the low-rank constraint contributes to the generalization ability,
which is essential for inferring the energy landscape
only from a small dataset. 
Thus, the multicomponent SPIM should serve as an efficient sampling machine to find
low-energy candidates using a trained low-rank Ising Hamiltonian.

Despite the importance of low-rank modeling in data science,
there has been no study on low-rank learning of Boltzmann machines,
to the best of our knowledge.
Our results
suggest the capabilities of the low-rank Boltzmann machine
as a statistical model with high parameter efficiency.
The low-rank learning may be further enhanced
by introducing sparsity regularization.
Elucidating the mechanism behind the gradient-ascent rule is also intriguing.

Another unique feature of the multicomponent SPIM is that
we can choose candidate states arbitrarily in MCMC algorithms
without any loss of the computation speed,
as demonstrated in multiple-spin flips for the knapsack problem.
Designing new MCMC algorithms
specialized for the multicomponent SPIM is considered important,
possibly by exploiting the low-rank property \cite{Koehler2022}
and physical noise \cite{Pierangeli2020NP}.

The high practical applicability of the multicomponent SPIM
highlights the need for hardware improvements in the SPIM,
such as enhanced computation speed and scalability, for further development.
It can exploit the low-rank property
in multiplexed architectures such as \cite{Luo2023}.
If the SPIM hardware allows $\bm\sigma$ to take
nonbinary, intermediate continuous values,
we can implement continuous-valued spin systems
with rich nonlinear dynamics
as in Refs.~\cite{Inagaki2016,Leleu2019,Goto2019,Suzuki2013SR,Suzuki2013PRE,Traversa2017,Bearden2020,Ercsey-Ravasz2011,Yamashita2020,Yamashita2021}.

In conclusion, we proposed the multicomponent computing model for the SPIM
that exhibits higher practical applicability
to various problems of combinatorial optimization and statistical learning
without losing the inherent scalability.
Notably, the proposed model has a unique affinity to
low-rank combinatorial optimization
and low-rank learning of Boltzmann machines.
These unexpected benefits of the SPIM architecture
are expected to contribute to the future development of non-von Neumann,
neuro-inspired computing.

\begin{acknowledgments}
This study was supported by JST CREST JPMJCR18K2.
H.Y. and H.S. appreciate the support from
JST Moonshot R\&D JPMJMS2021 and the WPI-IRCN at UTIAS.
\end{acknowledgments}

H.Y. and K.O. contributed equally to this work.

\bibliography{lowrank.bib}

\clearpage

\section*{Supplementary material}

# Supplemental material for

# Low-Rank Combinatorial Optimization and Statistical Learning by

# Spatial Photonic Ising Machine

Hiroshi Yamashita, Ken-ichi Okubo, Suguru Shimomura,

Yusuke Ogura, Jun Tanida, and Hideyuki Suzuki[*]

*Graduate School of Information Science and Technology,*

*Osaka University, Osaka 565–0871, Japan*

## I. OPTICAL SETUP FOR THE SPIM ARCHITECTURE

We implemented the SPIM architecture (Fig. 1 in the main text) for our proof-of-concept optical experiments. Light patterns with amplitudes $\xi_k$ are generated by using SLM1 (Santec, SLM-200, pixel size 1920 × 1080, pixel pitch: 8 $\mu$m) and a laser with a wavelength of 532 nm. Spin configuration $\sigma$ is encoded into the modulation pattern with binary phases by SLM2 (Hamamatsu Photonics, X15213-01, pixel size 1272 ×1024, pixel pitch: 12.5 $\mu$m). The light intensity $I_k$ is detected by a CMOS image sensor (Point Gray Research, Grasshopper GS3-U3-32S4, pixel pitch: 3.45 $\mu$m). Note that if $\xi_{i,k}$ is negative, we flip the spin $\sigma_i$ and use $|\xi_{i,k}|$ as the amplitude. Finally, we obtain the Ising Hamiltonian as a weighted sum of $I_k$ with coefficients $\lambda_k$ after capturing $I_k$ with different amplitudes $\xi_k$ for $K$ times.

## II. EXPERIMENTS ON A KNAPSACK PROBLEM

A knapsack problem generator [29] was used to obtain the following problem instance for the proof-of-concept experiment:

$$n = 13, \quad W = 80,$$
$$v = (7, 7, 8, 8, 2, 7, 12, 4, 9, 14, 2, 7, 14),$$
$$w = (6, 7, 1, 15, 14, 8, 5, 6, 4, 7, 5, 12, 10).$$

---

[*] hideyuki@ist.osaka-u.ac.jp

The optimal solution to this problem is $x = (1, 1, 1, 1, 0, 1, 1, 0, 1, 1, 1, 1, 1)$, for which the total value is 95 and the total weight is 80. The coefficients of the Ising formulation are $A = 2633$ and $B = 1$. The number of auxiliary variables is $m = 4$.

Note that the number of auxiliary variables for log trick is generally set as $m = \lfloor \log_2 W \rfloor + 1$; however, in this study, we set $m = \lceil \log_2 \max_i w_i \rceil$, assuming that the problem is nontrivial, $\sum_i w_i > W$. Coefficient $A$ for the constraint term has to be sufficiently large compared with $B$ for the total value term because the constraint violation should be penalized more than the gain of picking an item; specifically, we set $A = v_{\max} (2 \sum_i v_i - v_{\max}) + 1$ and $B = 1$, where $v_{\max} = \max_i v_i$.

In each of the optical experiment and numerical experiment, we generated 50 sample sequences of length 3000 using the Metropolis algorithm on the Gibbs distribution $P(\sigma) \propto \exp[-H(\sigma)/T]$ at a constant temperature $T = 10A$ for random initial spin configurations. In each time step, a candidate state $\sigma'$ is generated by flipping the spins of the current state $\sigma$ with a probability $3/(n + m)$ for each spin, except for the last one fixed to $\sigma_N = +1$.

## III. EXPERIMENTS ON MNIST DIGIT IMAGE DATA

For classification of handwritten digits, we prepared fully visible multicomponent models of size $N = 794$, composed of 784 units for $28 \times 28$ pixels of MNIST digit images and 10 units for the one-hot encoding of the 10 digit classes. We set the rank $K$ of the models to integer multiples of 10 in the range $10 \leq K \leq 790$ and trained each of them using the 60000 MNIST training images. To obtain the model expectation in the learning rule, we used the contrastive divergence (CD) algorithm [33]. We monitored the pseudo-likelihood and stopped training before the learning degraded. The hyperparameter values are common to all the rank values: the initial learning rates, $10^{-3}$ and $10^{-5}$ for $\lambda_k$ and $\xi_k$, respectively; the regularization coefficient, 0.001 for both $\lambda_k$ and $\xi_k$. Classification was performed by selecting the digit whose corresponding one-hot vector minimizes the Hamiltonian with the given image data. The classification accuracy was quantified as the ratio of the number of images classified accurately out of the 10000 MNIST test images.

For sampling digit images, we prepared 10 fully visible multicomponent models with size $N = 196$ and rank $K = 50$. We trained them using the MNIST training image data of each digit, in which the digit images were shrunk to $14 \times 14$ pixels and binarized. To obtain the model expectation in the learning rule, we used the persistent contrastive divergence (PCD) algorithm [36] with 100 persistent chains. The hyperparameter values are common to all digits: the initial

learning rates, $10^{-2}$ and $0.5 \times 10^{-3}$ for $\lambda_k$ and $\xi_k$, respectively; the regularization coefficient, 0.001 for both $\lambda_k$ and $\xi_k$. The reduced model was composed using only principal components with magnitudes $|\lambda_k| \|\xi_k\|^2 > 0.1$. Digit images were sampled by executing the Metropolis algorithm on each model for 1960 time steps with candidates generated by sequential single-spin flips starting from random initial spin configurations.

## IV. LEARNING RULE FOR THE MULTICOMPONENT MODEL

The learning rule for the multicomponent model is derived as the gradient-ascent algorithm on the log-likelihood, adopting the same approach as the ordinary Boltzmann machine.

For the model distribution with multicomponent Hamiltonian $H(\sigma)$,

$$P(\sigma) = \frac{1}{Z} \exp[-H(\sigma)], \quad Z = \sum_\sigma \exp[-H(\sigma)], \tag{S1}$$

the log-likelihood $\log L$ for a data distribution is given as follows:

$$\log L = \langle \log P(\sigma) \rangle_{\text{data}} = -\langle H(\sigma) \rangle_{\text{data}} - \log Z. \tag{S2}$$

Using the derivatives of $\log Z$ with respect to the model parameters,

$$\frac{\partial}{\partial \lambda_k} \log Z = \frac{1}{Z} \sum_\sigma \frac{\partial}{\partial \lambda_k} \exp[-H(\sigma)] = -\left\langle \frac{\partial H}{\partial \lambda_k} \right\rangle_{\text{model}}, \tag{S3}$$

$$\frac{\partial}{\partial \xi_k} \log Z = \frac{1}{Z} \sum_\sigma \frac{\partial}{\partial \xi_k} \exp[-H(\sigma)] = -\left\langle \frac{\partial H}{\partial \xi_k} \right\rangle_{\text{model}}, \tag{S4}$$

we obtain the derivatives of the log-likelihood as follows:

$$\frac{\partial}{\partial \lambda_k} \log L = -\left\langle \frac{\partial H}{\partial \lambda_k} \right\rangle_{\text{data}} + \left\langle \frac{\partial H}{\partial \lambda_k} \right\rangle_{\text{model}}, \tag{S5}$$

$$\frac{\partial}{\partial \xi_k} \log L = -\left\langle \frac{\partial H}{\partial \xi_k} \right\rangle_{\text{data}} + \left\langle \frac{\partial H}{\partial \xi_k} \right\rangle_{\text{model}}, \tag{S6}$$

which are identical to Eqs. (10) and (11) in the main text as the multicomponent Hamiltonian

$$H(\sigma) = -\frac{1}{2} \sum_k \lambda_k (\xi_k^\top \sigma)^2, \tag{S7}$$

has the following derivatives:

$$\frac{\partial H}{\partial \lambda_k} = -\frac{1}{2} \frac{\partial}{\partial \lambda_k} \sum_{k'} \lambda_{k'} (\xi_{k'}^\top \sigma)^2 = -\frac{1}{2} \xi_k^\top \sigma \sigma^\top \xi_k, \tag{S8}$$

$$\frac{\partial H}{\partial \xi_k} = -\frac{1}{2} \frac{\partial}{\partial \xi_k} \sum_{k'} \lambda_{k'} (\xi_{k'}^\top \sigma)^2 = -\lambda_k \sigma \sigma^\top \xi_k. \tag{S9}$$

We can introduce regularization in the learning process to avoid overfitting. We use the L2 regularizer on the parameters $\lambda_k$ and $\boldsymbol{\xi}_k$ in this study.



\end{document}